\documentstyle[preprint,pra,aps]{revtex}

\begin{document}

\draft

\title{Stability of Bose condensed atomic $^7$Li}

\author{M. Houbiers and H.T.C. Stoof}
\address{University of Utrecht, Institute for Theoretical
         Physics, Princetonplein 5, \\
         P.O. Box 80.006, 3508 TA  Utrecht, The Netherlands} 
\maketitle

\begin{abstract}
We study the stability of a Bose condensate of atomic $^7$Li in
a (harmonic oscillator) magnetic trap at non-zero temperatures. 
In analogy to the stability criterion for a neutron star, 
we conjecture that the gas becomes 
unstable if the free energy as a function of the central density
of the cloud has a local extremum which conserves the
number of particles. Moreover, we show that 
the number of condensate particles at the point of instability decreases 
with increasing temperature, and that 
for the temperature interval considered, the normal
part of the gas is stable against density fluctuations
at this point.
\end{abstract}


\section{INTRODUCTION}
\label{inleiding}
Since several groups reported evidence for Bose-Einstein condensation
(BEC) in atomic gas samples of $^{87}$Rb \cite{JILA},
$^{23}$Na \cite{MIT}\ and possibly $^7$Li \cite{rice} last year, 
there has been an increased interest in this field of physics.
After almost two decades, one has finally been able to verify 
the prediction which Bose and in particular 
Einstein launched for
so many years ago. This opens a totally new era regarding research on
degenerate systems, and it is to be expected that there
will appear many new and interesting experimental and theoretical
results about the properties of Bose condensed atomic gases 
in the near future. 

This is certainly true for the $^{87}$Rb and the $^{23}$Na systems, but
maybe less so for the $^7$Li system, because the experimental results
on the latter gas are not yet completely
understood. In contrast to the former systems where the atoms have a
positive s-wave scattering length $a$, $^7$Li atoms have a negative 
s-wave scattering length. This leads to an effectively attractive
interatomic interaction, which makes the system unstable at large densities.
Indeed one of us showed that a dilute, 
homogeneous gas of atoms with a negative s-wave scattering length $a$
collapses to a dense (liquid or solid) state 
before the density is reached at which BEC is expected 
to occur at a given temperature \cite{henk}. Furthermore, it was shown that the
BEC transition is actually preceded by a BCS-like
transition to a superfluid state and that
this transition also occurs in the unstable regime of the
phase diagram. 

However, as was pointed out by Hulet \cite{randy2}\
and Ruprecht {\it et al.} \cite{keith} the situation is different if the
atoms are confined by a magnetic trap. In particular,
it was found that for an inhomogeneous
gas cloud at zero temperature the condensate is stable at the mean-field
level if the number of particles is sufficiently small, or more precisely
if $N < N_{0,max} \equiv 0.573 l/|a|$,
where $l=\sqrt{\hbar/m\omega}$\ is the typical size of the one-particle
ground state in the trap. 
However, quantum fluctuations cause a decay of the condensate 
on a timescale which is 
fortunately much longer than the timescales at which
the experiments were performed. The same holds true at the relatively high
temperatures of interest experimentally, even though the decay is now caused
by thermal fluctuations \cite{stoof}. 

However, in Ref.~\cite{stoof}\ only the stability of the condensate was 
discussed. Although this is an important first step, it is clearly not
sufficient, because we know from the homogeneous case that also the
non-condensed gas can be unstable against density fluctuations.
Therefore we will consider here the stability of the complete system including
condensate as well as above condensate particles. We will try
to answer the question when the system as a whole 
becomes unstable and whether it is the condensate part or the non-condensate
part which causes the instability.
 
In this stability analysis we cannot make use of the simple local-density
approximation. The local-density approximation 
is only applicable in systems for which the correlation length $\xi$\ 
(roughly speaking the distance over which the particles influence each other) 
is much smaller
than the typical trap size $l $\ over which
the density changes and the system behaves locally homogeneous.
However, close to the critical temperature the correlation length 
diverges, and the local-density approximation always breaks down.
Nevertheless, it is valid at the spinodal point if we satisfy the 
condition that 
$k_B T/\hbar \omega \gg l/|a| \gg 1$ or equivalently $N \gg N^3_{0,max}$ 
\cite{stoof,michel}, i.e.\
the total number $N$\ of particles in the gas must be much
larger then the third power of the maximum number of condensate
particles $N_{0,max}$. Since this amounts
to $N \gg 10^9$\ for the trap parameters of the Rice experiment \cite{rice}, 
it is clear that on the basis of the local-density approximation
we cannot decide if a cloud of $10^4-10^5$\ atoms is mechanically
stable and therefore (meta)stable. 
On the other hand, if it were allowed to use the local density approximation
at the spinodal point, 
we would immediately conclude that a (meta)stable condensate cannot exist.

To go beyond the local-density approximation, we will present
numerical results for the free energy
of the system at several temperatures.
To do so, we first present in Sec.~\ref{theorie}\ the finite
temperature theory for the inhomogeneous gas. The equations of motion that
describe the gas are derived from a variational principle 
\cite{michel}. In analogy to the homogeneous case, we incorporate the
possibility of both a BEC and a BCS transition. 
Subsequently we give an expression
for the free energy. Since the experiments with $^7$Li are performed
at densities and temperatures such that $na\Lambda_{th}^2 \ll 1$
(where $n$\ is the density and 
$\Lambda_{th}=\sqrt{2\pi\hbar^2/mk_BT}$\ the thermal
wavelength of the atoms) and in particular even at $T\simeq T_c$, the theory
can be simplified by neglecting the possible BCS pairing and 
using the Hartree-Fock approximation. 
However, in view of the fast experimental developments, it is to be expected
that also the regime $na\Lambda_{th}^2 \gg 1$, which amounts to 
$T\ll T_c$, can be reached in the
near future and this is why we present here the full theory including
also the effects of BCS pairing. 
In Sec.~\ref{resultaat}\ we then present our numerical results.
In Sec.~\ref{Tnul}\ we discuss 
the zero temperature limit of the Hartree-Fock
approximation and make a
comparison with previous calculations. Also a thermodynamic
criterion for the stability of the gas is given.
In Sec.~\ref{Teindig}\ we proceed to non-zero temperatures.
We calculate the maximum number of condensate particles as a function of
temperature and pronounce upon the issue whether the condensate or
the non-condensate causes the instability of the gas.
The paper ends in Sec.~\ref{conclusie}\ with some conclusions. 

\section{THEORY}
\label{theorie}
We start this paper with the equilibrium theory 
for a dilute gas of particles with mass $m$\ in 
an external trap potential $V_{ext}(\vec{r})$,
interacting
with each other through an approximately local (because $|a| \ll l$)
two-body potential $V_0 \delta(\vec{r})$. 
In the numerical calculations
which follow subsequently, we will specialize to $^7$Li atoms, which have
a negative s-wave scattering length $a$\ and $V_0 < 0$. 
The interparticle interaction is therefore effectively attractive.

The grand-canonical Hamiltonian of the system is given by \cite{FenW} 
\begin{equation}
H =  \int d\vec{r} \left\{ \psi^{\dag}(\vec{r}) \left(
-\frac{\hbar^2 \nabla^2}{2m} + V_{ext}(\vec{r}) - \mu
 \right) \psi(\vec{r}) + \frac{1}{2} V_0 \psi^{\dag}(\vec{r}) \psi^{\dag}
(\vec{r}) \psi(\vec{r}) \psi(\vec{r}) \right\},
\end{equation}
where $\mu$\ is the chemical potential and 
$\psi(\vec{r})$\ respectively $\psi^{\dag}(\vec{r})$\ annihilates and creates a
particle at position $\vec{r}$. 
As usual, the density of particles in the system $n(\vec{r})$\ is given by
the grand canonical average
$\langle \psi^{\dag}(\vec{r}) \psi(\vec{r}) \rangle$, and the
total number of particles is $N= \int d\vec{r} \ n(\vec{r})$, which
ultimately determines the chemical potential of the gas.

For particles with a positive s-wave scattering length $a$, the annihilation
operator $\psi(\vec{r})$\ has a non-vanishing expectation value
below the critical temperature. By
separating out this expectation value in the usual way \cite{FenW}, i.e.
\begin{equation}
\psi(\vec{r}) = \psi_0(\vec{r}) + \psi'(\vec{r}),
\end{equation}
where $\psi'$\ describes the non-condensate part and $\psi_0(\vec{r}) = \langle
\psi(\vec{r}) \rangle = \sqrt{n_0(\vec{r})}$\ is the condensate wave
function, one can derive the equations of motion for the condensate
as well as the non-condensate part of the gas. However, in the case of
a negative s-wave scattering length, it can be shown for the homogeneous case 
that the free energy as a function of
the expectation value of the field operator $\psi$\ has at low temperatures
a local {\it maximum} for some non-zero value of $\langle \psi(\vec{r})
\rangle$\ and there is only a local {\it minimum} for $\langle
\psi(\vec{r}) \rangle = 0$, which is therefore the 
correct value around which one has to expand \cite{michel}. 
As a result we must use a different order parameter to describe a
phase transition due to quantum degeneracy effects, namely 
the BCS-type order parameter $\langle \psi(\vec{r}) \psi(\vec{r}) \rangle$. 
In the case of bosons, this is actually known as the 
Evans-Rashid order parameter. 

\subsection{Evans-Rashid transition}
To derive the equations of motion that describe the gas,
it is useful to determine the (exact) grand-canonical
potential
\begin{equation}
\Omega_{ex}(T,\mu) = - k_B T \ln{(\mbox{Tr} \exp{(-\beta H )})},
\end{equation}
from which all thermodynamic quantities that we wish to know
can be calculated. It is well known that if $H_t$\ is some
trial Hamiltonian, and $\Omega_t$\ the corresponding grand-canonical
potential, we have the variational principle \cite{falk}
\begin{equation}
\Omega_{ex} \leq \Omega \equiv \Omega_t + \langle H - H_t \rangle_t
\label{omega}
\end{equation}
where $\langle O \rangle_t$\ is the expectation value of the operator $O$\
in the grand-canonical ensemble based on $H_t$. 
The trial Hamiltonian that we want to use here is given by
\begin{eqnarray}
H_t & = & \left. \int d\vec{r} \right\{ \psi^{\dag}(\vec{r}) \left(
-\frac{\hbar^2 \nabla^2}{2m} + V_{ext}(\vec{r})  
- \mu  + \hbar\Sigma(\vec{r}) \right)\psi(\vec{r}) + \nonumber \\
 & &+ \frac{1}{2} \Delta_0(\vec{r}) \psi^{\dag}(\vec{r}) \psi^{\dag}
(\vec{r}) + \left.\frac{1}{2} \Delta_0^* (\vec{r}) \psi(\vec{r}) \psi(\vec{r})
\right\}. 
\label{nondiag}
\end{eqnarray}
It is quadratic in the field operators and 
indeed has non-zero expectation values for both
$\langle \psi^{\dag}(\vec{r}) \psi(\vec{r}) \rangle_t$\ and $\langle
\psi(\vec{r}) \psi(\vec{r}) \rangle_t$. 
In this expression, the functions $\hbar\Sigma(\vec{r})$, $\Delta_0(\vec{r})$\
and it's complex conjugate $\Delta_0^*(\vec{r})$\ are variational parameters
which have to be determined by minimization of the grand-canonical
potential $\Omega$. The trial Hamiltonian has non-diagonal elements proportional
to the BCS order parameter $\Delta_0(\vec{r})$\ and the diagonal contribution
proportional to $\hbar\Sigma(\vec{r})$\ is the self-energy due to the
two-body interaction. 

This trial Hamiltonian can be put into the diagonal form
\begin{equation}
H_t = E_g + \sum_j \ \hbar\omega_j b^{\dag}_j b_j
\label{diag}
\end{equation}
by applying the Bogoliubov transformation
\begin{mathletters}
\begin{equation}
\psi(\vec{r})  =  \sum_j \left( u_j(\vec{r}) b_j + v^*_j(\vec{r}) b^{\dag}_j
\right)  \label{trafo1} 
\end{equation}
\begin{equation}
\psi^{\dag}(\vec{r})  = 
\sum_j \left( v_j(\vec{r}) b_j + u^*_j(\vec{r}) b^{\dag}_j
\right).
\label{trafo2}
\end{equation}
\label{trafo}
\end{mathletters}
The operators $b_j$\ and $b^{\dag}_j$\ are required to
satisfy the usual Bose commutation relations,
and therefore the functions $u_j(\vec{r})$\ and $v_{j}(\vec{r})$\ are 
normalized as
\begin{equation}
\left[ \psi(\vec{r}), \psi^{\dag}(\vec{r}') \right] =
\sum_j \left(u_j(\vec{r}) u_j^*(\vec{r}') - v_{j}^* (
\vec{r}) v_j(\vec{r}') \right) = \delta( \vec{r} - \vec{r}' ).
\label{normering}
\end{equation}
Using the relations
$\left[ H_t, b_j \right] = - \hbar \omega_j b_j$\
and $\left[ H_t, b^{\dag}_j \right] =  \hbar\omega_j b^{\dag}_j$, 
and substituting Eqs.~(\ref{trafo1}) and (\ref{trafo2})\ into the commutators 
$\left[ H_t, \psi \right] $\
and $\left[ H_t, \psi^{\dag} \right] $, 
it is found that $u_j(\vec{r})$\ and $v_j(\vec{r})$\ must be solutions to
the following eigenvalue equation:
\begin{equation} 
\left( \begin{array}{cc}
H_0  + \hbar \Sigma(\vec{r})  - \mu  - \hbar \omega_j  &
\Delta_0(\vec{r}) \\
-\Delta_0^*(\vec{r}) & -  H_0 -
\hbar \Sigma(\vec{r}) + \mu -\hbar \omega_j  
\end{array}
\right) \cdot
\left( \begin{array}{c} u_j(\vec{r}) \\ v_j(\vec{r}) \end{array} \right) =0, 
\label{matrix}
\end{equation}
where $H_0 = -\hbar^2 \nabla^2/2m + V_{ext}(\vec{r}) $. 
These coupled equations are the Bogoliubov-DeGennes equations
\cite{degennes}. They can 
be solved selfconsistently once the functions $\hbar \Sigma(\vec{r})$\ and
$\Delta_0(\vec{r})$ are obtained from minimization of $\Omega$ and
expressed in terms of $u_j(\vec{r})$\ and $v_j(\vec{r})$. Furthermore 
the ground-state energy $E_g$\ in Eq.~(\ref{diag}) is given by
\begin{equation}
E_g = \int d\vec{r} \sum_j \ \left\{ \ -\hbar\omega_j |v_j(\vec{r})|^2
-\frac{1}{2} \Delta_0 u_j^*(\vec{r}) v_j(\vec{r}) + \frac{1}{2}
\Delta_0^*(\vec{r}) u_j(\vec{r}) v_j^*(\vec{r}) \right\}.
\label{Egrond}
\end{equation} 

We now return to the calculation of the thermodynamic potential $\Omega$.
Since $H_t$\ is diagonal according to Eq.~(\ref{diag}), it is easily
verified that the first term on the right-hand side
of Eq.~(\ref{omega}) is given by
\begin{equation}
\Omega_t = E_g + k_B T \sum_{j} \ln{\left(1- e^{-\beta\hbar\omega_j }
\right) },
\end{equation}
whereas the second term can be rewritten, using Wick's theorem
\cite{FenW}, as
\begin{eqnarray}
\langle H-H_t \rangle_t & = & \int d \vec{r}  \left\{
V_0 \langle \psi^{\dag}(\vec{r}) \psi(\vec{r}) \rangle_t \langle
\psi^{\dag}(\vec{r}) \psi(\vec{r}) \rangle_t + 
\frac{1}{2} V_0 \langle \psi^{\dag}(\vec{r}) \psi^{\dag}(\vec{r})
\rangle_t \langle\psi(\vec{r})\psi(\vec{r})\rangle_t \right. \nonumber \\ 
 & & \left. - \hbar\Sigma(\vec{r})\langle\psi^{\dag}(\vec{r})
\psi(\vec{r}) \rangle_t
- \frac{1}{2} \Delta_0(\vec{r}) \langle \psi^{\dag}(\vec{r}) 
\psi^{\dag}(\vec{r}) \rangle_t -
\frac{1}{2} \Delta_0^*(\vec{r}) \langle \psi(\vec{r}) 
\psi(\vec{r}) \rangle_t \right\}.
\end{eqnarray}
Substituting the Bogoliubov transformation from Eq.~(\ref{trafo}),
we find that 
\begin{equation}
\langle \psi^{\dag}(\vec{r}) \psi(\vec{r}) \rangle_t   = 
 \sum_j \left\{ (|u_j(\vec{r})|^2 + |v_j(\vec{r})|^2)
N(\hbar \omega_j) + |v_j(\vec{r})|^2 \right\}, 
\label{exp1} 
\end{equation}
and
\begin{equation}
\langle \psi(\vec{r}) \psi(\vec{r}) \rangle_t   = 
\sum_j u_j(\vec{r}) v_j^*(\vec{r}) ( 1 +2 N(\hbar \omega_j) ),
\label{exp2}  
\end{equation}
where the function $N(\hbar \omega_j) 
= \langle b^{\dag}_j b_j \rangle_t = (e^{\beta \hbar \omega_j} -1) ^{-1}$\ is
the Bose distribution function for the Bogoliubov quasiparticles.  

The still unknown functions $\hbar \Sigma(\vec{r})$, 
$\Delta_0^*(\vec{r})$\ and  
$\Delta_0(\vec{r})$\ have to be chosen such that the functional
$\Omega[\hbar\Sigma,\Delta_0,\Delta_0^*]$\ is minimal, i.e.
\begin{equation}
\left. \frac{\partial \Omega}{\partial \hbar\Sigma} \right|_{
\Delta_0,\Delta_0^*} = 
\left. \frac{\partial \Omega}{\partial \Delta_0} \right|_{
\hbar\Sigma,\Delta_0^*} = 
\left. \frac{\partial \Omega}{\partial \Delta_0^*} \right|_{
\hbar\Sigma,\Delta_0} =  0.
\label{min}
\end{equation}
The last condition is just the complex conjugate of the second, and it
suffices to consider only one of them.  
In the Appendix it is shown that Eq.~(\ref{min}) requires that  
\begin{equation}
\hbar\Sigma(\vec{r}) = 2 V_0 \langle \psi^{\dag}(\vec{r}) \psi(\vec{r})
\rangle_t
\label{sigma}
\end{equation}
and
\begin{equation}
\Delta_0(\vec{r}) = V_0 \langle \psi(\vec{r}) \psi(\vec{r}) \rangle_t.
\label{gap}
\end{equation}
As is explained in Ref.~\cite{henk2}, to incorporate all two-body 
scattering processes in this many-particle system,
the factor $V_0$\ in Eq.~(\ref{sigma}) must be replaced by the many-body
T-matrix $T^{MB}$, but in the regime of interest where the temperature
is large compared to the average interaction energy, i.e.\  
$na\Lambda_{th}^2 \ll 1$, this can be approximated by the two-body
scattering matrix $T^{2B} = 4\pi a \hbar^2/m$\ and we find the usual 
Hartree-Fock contribution to the self-energy $\hbar \Sigma(\vec{r})
= 2 n(r) T^{2B}$. In addition,
Eq.~(\ref{gap}) corresponds to the gap equation of BCS theory.
Since this theory already 
incorporates all ladder diagrams, the factor $V_0$\ should 
here not be replaced by the many-body T-matrix. 
Collecting together all terms, we find for the thermodynamic potential
\begin{equation}
\Omega =  E_g + k_B T \sum_j \ln(1- e^{- \beta \hbar \omega_j})
-  \int d\vec{r} \left[ n^2(\vec{r}) T^{2B} - \frac{1}{2} \frac{
|\Delta_0(\vec{r})|^2}{V_0}  \right].
\label{om}
\end{equation}
From this expression the free energy of the system can be calculated directly
using the thermodynamic identity $F= \Omega + \mu N$.

The equations obtained thus far are only 
valid when there is no Bose condensate
present. However, as in the homogeneous case, it is evident that
the lowest energy $\hbar \omega_0$\ will go through zero 
at sufficiently low temperatures and at this point 
the corresponding one-particle
ground state becomes macroscopically occupied, i.e.\ a Bose condensate
is formed. Hence, this ground state has then to be considered explicitly.

\subsection{Bose-Einstein condensation}
\label{bec}

We now adress the changes in the above equations that
are required if a Bose condensate is present. 
First, we consider the limit $T\rightarrow 0$, for which all particles
in the system tend to occupy the ground state. In the (trial) 
grand-canonical ensemble
\begin{equation}
Z_{t} = \mbox{Tr} [ \exp{(-\beta H_t )} ] = \mbox{Tr} [ 
\exp{(-\beta E_g -\beta \sum_j \hbar \omega_j b_j^{\dag} b_j)} 
\end{equation}
used thus far, one can calculate by standard-statistical physics
methods, that for
any $j$\ the expectation value 
$\langle b_j^{\dag} b_j^{\dag} b_j b_j \rangle_t = 
\langle N_j^2 \rangle_t - \langle N_j \rangle_t = 2\langle N_j \rangle_t^2$,
where $\langle N_j \rangle_t = -(1/ \beta) 
\partial \ln{Z_t} / \partial \hbar \omega_j = N(\hbar \omega_j)$. 
In fact, the factor of 2 in the self-energy
Eq.~(\ref{sigma}) originates from this property. However, at zero 
temperature all particles will tend to the ground state and we therefore
find that the fluctuations in the total number of particles are given by 
\begin{equation}
\frac{ \langle N^2 \rangle - \langle N \rangle^2}{\langle N \rangle^2}
= \frac{ \langle N_0^2 \rangle - 
\langle N_0 \rangle^2}{\langle N_0 \rangle^2} \simeq 1 + \frac{1}{\langle
N_0 \rangle} = 1 + \frac{1}{\langle N \rangle}
\end{equation}
and of order ${\cal{O}}(1)$\ 
instead of the usual ${\cal{O}}(1 / \sqrt{\langle N \rangle})$. Hence 
the fluctuations in the particle number
are as large as the average itself, which leads to the conclusion that
the use of the grand-canonical ensemble does not lead to an 
appropriate description of the Bose condensed gas. 
 
There are basically two ways to restrict the grand-canonical ensemble and 
circumvent this problem. The first
one is to introduce a condensate expectation value according to
$\langle b_0 \rangle = \sqrt{N_0}$. Clearly, 
we then have $\langle N_0^2 \rangle = N_0^2$, and at zero temperature
we end up with $\hbar \Sigma = n_0 T^{2B}$\ instead of Eq.~(\ref{sigma}).
For non-zero temperatures the effect will be that the ground-state wave
function satisfies Eq.~(\ref{matrix}) with $\hbar \Sigma = (2n -n_0)T^{2B}$,
whereas the excited state wave functions obey 
the same equation except that now $\hbar \Sigma = 2 n T^{2B}$. 
Moreover, the free-energy density now 
contains the term $f= \frac{1}{2} n_0^2 T^{2B}$\ instead of 
$n_0^2 T^{2B}$\ (c.f. Eq.~(\ref{om}) and use $f=\Omega/V + \mu n$). 
Thus from energy considerations,
it is indeed favorable to introduce a condensate expectation value
if the s-wave scattering length is positive ($T^{2B} > 0$), but this is 
not the case for negative $a$. As mentioned previously, it can be shown that
the thermodynamic potential $\Omega(\langle \psi \rangle )$\ is a
mexican hat shaped function with extremum at
$| \langle b_0 \rangle | = \sqrt{N_0}$, which is however inverted with
respect to the $\langle b_0 \rangle$-plane when the scattering length
$a$\ changes from positive to negative \cite{michel}. 
Therefore, the local minimum at $\langle b_0 \rangle
= \sqrt{N_0}$\ that is present for $a > 0$\ becomes a local maximum
for $a< 0$, and the local minimum of the thermodynamic potential occurs at 
$\langle b_0 \rangle =0$\ in
the latter case. So, for a gas of $^7$Li atoms, the use of the 
order parameter $\langle b_0 \rangle$\ appears not to be the correct
way to control the fluctuations in the number of condensate particles. 

The second method to restrict the condensate fluctuations 
is to introduce a different restricted grand-canonical ensemble 
according to
\begin{eqnarray}
Z_{t} &= &\mbox{Tr} \left[ \mbox{e}^{-\beta H_t } \right] \nonumber \\
& = & \sum_{N_0} \mbox{Tr} \left[ \mbox{e}^{ - \beta E_g
-\beta \sum_j \hbar \omega_j 
b^{\dag}_j b_j }\ \delta_{b^{\dag}_0 b_0 , N_0} \right] \nonumber \\
& \equiv & \sum_{N_0} \mbox{e}^{-\beta \Omega_t(\mu,N_0)}
\label{zgroot}
\end{eqnarray}
where
\[
\mbox{e}^{-\beta \Omega_t(\mu,N_0)} = \mbox{e}^{-
\beta E_g - \beta \hbar\omega_0 N_0}
\prod_{j \neq 0} \exp{(-\ln{(1-\mbox{e}^{-\beta \hbar \omega_j })})}.
\]
From these expressions it follows that
\[
\Omega_t(\mu, N_0) = E_g + \hbar \omega_0  N_0 + k_BT \sum_{j\neq 0}
\ln{(1- \mbox{e}^{-\beta \hbar \omega_j })}.
\]
The total number of particles is therefore given by $N= 
\sum_j \langle N_j \rangle_t = \sum_j
\partial \Omega_t /  \partial \hbar \omega_j = 
N_0 + \sum_{j\neq 0} N(\hbar \omega_j)$.
Moreover, the largest contribution to the sum over the number of condensate 
particles in Eq.~(\ref{zgroot}) comes from a minimum in $\Omega_t$,
which implies that $N_0 =0$\ if $\hbar \omega_0 < 0$\ and 
$0 < N_0 < \infty$\ if $\hbar \omega_0 =0$.

The expectation value $\langle N_0^2 \rangle$\ calculated in 
this restricted grand-canonical ensemble (by construction)
equals $\langle N_0 \rangle^2$, whereas for the other energy
levels ($j \neq 0$) nothing has changed compared with the results in 
the original grand-canonical ensemble. In conclusion we therefore arrive
at the {\it same} equations for the ground state and the
excited state wave functions 
as we had derived by the first method for a gas with positive
s-wave scattering length:  
The ground-state
wave function $(u_0(\vec{r}),\ v_0(\vec{r}))$\ satisfies Eq.~(\ref{matrix}) 
with $\hbar \Sigma = \left[ 2n(\vec{r}) - n_0(\vec{r}) \right] T^{2B}$\
whereas the excited states have just $\hbar \Sigma = 2 n(\vec{r}) T^{2B}$.
In addition, the condensate density obeys
$n_0(\vec{r}) = N_0 ( |u_0(\vec{r})|^2 + |v_0(\vec{r})|^2 ) + 
|v_0(\vec{r})|^2$\ and the total density is given
by $n(\vec{r}) = n_0(\vec{r}) +
\sum_{j\neq 0} \{ N(\hbar \omega_j)(|u_j(\vec{r})|^2 +
|v_j(\vec{r})|^2) + |v_j(\vec{r})|^2$ \}. 
The change in expectation value $\langle N_j^2 \rangle$\ for
$j=0$\ will of course also change the
free energy if the system is Bose condensed. Taking this change into
account, the grand-canonical potential $\Omega$\ turns out 
to be given by Eq.~(\ref{om})
except that the term $j=0$\ must be excluded from the summation over states 
and a term $\frac{1}{2} \int d\vec{r}\ n_0^2(\vec{r}) T^{2B}$ must be
added. As a result, the free energy becomes 
\begin{equation}
F= \mu \int d\vec{r}\ n(\vec{r})\ + E_g + k_BT\sum_{j\neq 0}
\ln{(1- e^{-\beta \hbar \omega_j})} \ - \int d\vec{r} \left[
(n^2(\vec{r}) - \frac{1}{2} n_0^2(\vec{r})) T^{2B} -\frac{1}{2}
\frac{|\Delta_0(\vec{r})|^2}{V_0} \right].
\label{free10}
\end{equation} 
Note that the self energy $\hbar \Sigma$\ for the groundstate contains
a term $n_0(\vec{r}) T^{2B}$. According to Ref.~\cite{henk2}, the two-body
scattering matrix in this term should again have been the many-body T-matrix.
However, at sufficiently high temperatures such that $na\Lambda^2_{th} \ll 1$,
or even in the opposite regime if the interaction energy 
between the atoms is smaller than
the energy splitting of the one-particle states, i.e.\ $n T^{2B} < \hbar \omega$,
the many-body T-matrix $T^{MB}$ can be approximated by $T^{2B}$.  

\subsection{Mechanical stability}
\label{stab}

For the homogeneous gas with effectively attractive interactions, 
the free-energy density satisfies
$\partial f / \partial n = \mu$, and the chemical
potential $\mu$\ as a function of the density $n$\ becomes multivalued
($\partial^2 f / \partial n^2$\ changes sign)
for smaller densities than those needed for BEC. This is the 
instability criterium for the homogeneous system. A detailed analysis
shows that in this case the BEC transition is indeed preceded by
a BCS transition, but that both transitions occur in the unstable
regime of the phase diagram. 

However, in the introduction we already mentioned that
in an inhomogeneous system a metastable
condensate can exist if the number of condensate particles
is sufficiently small, i.e. $N_0 < 0.573 l/|a|$ \cite{keith}.
Qualitatively this can be understood from the fact that  
a collapse of the condensate requires that other harmonic oscillator
states need to be mixed into the wave function of the condensate.
For this energy is needed (virtually), which can be 
supplied by the interactions 
provided that the densities in the system are sufficiently high. 
As a result, when the density
of the gas becomes so high that the system will collapse, there
must be some radial unstable mode in the density fluctuations. 

Systems of compact objects such as
white dwarfs and neutron stars \cite{shapiro} can also be unstable
for collapse under certain conditions. Although a compact object consists of
degenerate fermions (electrons, protons and neutrons), 
and furthermore time and length
scales for stellar systems cannot be compared with those of the Bose
condensate, the physics in both systems has interesting similarities.

Indeed, the final state of a star when it has burnt up all it's nuclear
fuel is a white dwarf, neutron star or black hole,
depending on it's mass. Such a compact object is formed because 
in the final stages the star still radiates energy 
at the expense of
gravitational energy, i.e.\ the system contracts. This cannot go on
indefinitely, because at a certain point, the electrons
and protons in the star become degenerate. This causes an extra internal 
pressure and the star will come to equilibrium. For a white
dwarf, which has a maximum mass of 1.4 times the solar mass,
this occurs at a radius of about 5000 km. In this stage the gradient of
the pressure just cancels the gradient in potential energy.   
However, when the mass of the original star is between 1.4 and 3
times the solar mass,
the gravitational force will be so strong that equilibrium can 
only be restored when almost all electrons and protons are
squeezed together to neutrons by inverse $\beta$-decay: In that case the 
star contracts to an even more compact neutron star with a radius of about 10 km.  
Above these definite maximum masses, the white dwarf respectively
the neutron star can not support themselves against gravitational
collapse and this can lead to black hole formation.
To study the stability of these systems, it is known that it is convenient to
parametrize all equilibrium density profiles $n(\vec{r};n_c)$ by the
central density $n_c$\ of the star and that at the point of 
instability the mass of the object as a function of the central 
density of matter $n_c$\ exhibits an extremum. 

In analogy, we thus expect that in the case of a trapped
atomic gas, the onset of the instability is determined by the 
condition $\partial F /\partial n_c = 0$\ and that there exists
a zero mode in the density fluctuations at this point. 
To see this more explicitly, we consider the free energy functional
$F[n]$, which gives the free energy of the equilibrium density
profile $n(\vec{r};n_c)$. As a result we have
\begin{eqnarray}
F[n(\vec{r};n_c) + \delta n(\vec{r})] & = &F[n(\vec{r};n_c)] +
\int d\vec{r} \ \mu (n_c) \delta n(\vec{r}) +  \nonumber 
\\ & & + \frac{1}{2} \int d\vec{r} d\vec{r}' \ \left. 
\frac{\delta^2 F}{\delta n(\vec{r}) \delta
n(\vec{r}') } \right|_{n(\vec{r};n_c)} \ 
\delta n(\vec{r}) \delta n(\vec{r}') +.... 
\end{eqnarray}
where $\mu(n_c) = \left. \delta F / \delta 
n(\vec{r}) \right|_{n(\vec{r};n_c)}$.
If the central density is changed slightly, we have
\begin{eqnarray}
F[n(\vec{r};n_c + \delta n_c)] & =  & F[n(\vec{r};n_c) +
\frac{\partial n(\vec{r}; n_c)}{\partial n_c} \delta n_c 
+ {\cal O}(\delta n_c^2) ] \nonumber \\ & = &
F[n(\vec{r};n_c)] + \int d \vec{r}\ \mu(n_c) \frac{\partial
n(\vec{r};n_c)}{\partial n_c} \delta n_c + {\cal O}(\delta n_c^2).
\label{DFDN}
\end{eqnarray}
We thus conclude that if $\partial F / \partial
n_c = 0$, then either $\mu(n_c) =0$\ or $\int d\vec{r}\ \partial
n(\vec{r};n_c) / \partial n_c = 0$. The latter possibility 
is in general the physically relevant one because it shows that the two
density profiles $n(\vec{r};n_c)$\ and $n(\vec{r};n_c + \delta n_c)$\
have the same total number of particles, i.e. 
\begin{equation}
N(n_c + \delta n_c) = \int d\vec{r}\ n(\vec{r}; n_c + \delta n_c)
\simeq \int d\vec{r}\ n(\vec{r};n_c) = N(n_c).
\end{equation}
The fact that two density profiles containing the same number of particles 
have the same free energy up
to first order indicates that there is a zero mode present. 
Roughly speaking, it does not
cost energy to deform the first density profile continuously
into the second, which indicates the threshold for instability. 
We therefore anticipate that the onset of instability occurs
if the free energy has an extremum which conserves particle number 
(i.e.\ $\mu (n_c) \neq 0$)
as a function of the central density
of the gas. So, although collapse in compact objects respectively
in a Bose condensate is caused by a different mechanism, the final
criterion in both system may be, surprisingly enough, the same. 

\subsection{Hartree-Fock approximation}

We have derived the equations that describe an inhomogeneous gas 
at non-zero temperatures. A convenient procedure to solve these 
equations numerically, would be to start with some suitable 
initial distribution of particles $n(\vec{r})$\ and
an initial BCS order parameter $\Delta_0(\vec{r})$, and iterate the equations
to selfconsistency. It is, however, well known that it is rather
difficult to ensure the selfconsistency  
of $\Delta_0(\vec{r})$\ \cite{degennes}, and furthermore that Eq.~(\ref{gap})
contains a divergence, because the interparticle potential was approximated
by a $\delta$-function potential. For a homogeneous gas, this divergence
can easily be corrected for, but in this inhomogeneous case, it is not {\it a
priori}\ clear how one has to deal with it properly, although
it is not difficult to convince oneself that the divergence can be
canceled by calculating the molecular states of two atoms in the trap. 
Fortunately it is not necessary to solve these problems here because we
are primarily interested in the regime $n a \Lambda_{th}^2 \ll 1$, where 
the average energy $k_B T$\ of the particles is much larger than the
interaction energy and the effect of $\Delta_0(\vec{r})$\ is very small.  
Therefore, we neglect in the following 
the BCS order parameter $\Delta_0(\vec{r})$, which in turn means that
the functions $v_j(\vec{r}) = 0$. 
So, for the uncondensed gas, the Bogoliubov-DeGennes equation 
Eq.~(\ref{matrix}) then reduces to the 
Schr\"odinger equation for a particle in an effective
potential $V_{eff}(\vec{r}) = V_{ext}(\vec{r}) + 2 n(\vec{r}) T^{2B}$
\begin{equation}
\left( -\frac{\hbar^2 \nabla^2}{2m} + V_{ext}(\vec{r}) 
+ 2 n(\vec{r}) T^{2B} - \mu \right)
\phi_j(\vec{r}) = \hbar \omega_j \phi_j(\vec{r}).
\label{schrodinger}
\end{equation}
and the free energy of the system is
\begin{equation}
F = \Omega + \mu N =k_B T \sum_j \ln {\left( 1 - \exp{(-\beta
\hbar \omega_j ) } \right) } - \int d\vec{r}\ n^2(\vec{r})T^{2B} 
+ \mu \int d\vec{r}\ n(\vec{r}),
\label{vrijeenergie}
\end{equation}
where the particle density is given by 
\begin{equation}
n(\vec{r}) = \sum_j |\phi_j(\vec{r})|^2 N(\hbar \omega_j),
\end{equation}
and the wave functions $\phi_j(\vec{r})$\ are subject to the condition 
\begin{equation}
\int d \vec{r} \ |\phi_j(\vec{r})|^2 = 1.
\end{equation}

When the ground state is macroscopically occupied, the noncondensed 
particles satisfy Eq.~(\ref{schrodinger})
for $j \neq 0$, and the condensate wave function satisfies
\begin{equation}
\left( -\frac{\hbar^2 \nabla^2}{2m} + V_{ext}(\vec{r}) 
+ [ 2 n(\vec{r})-n_0(\vec{r})] T^{2B} - \mu \right)
\phi_0(\vec{r}) = \hbar \omega_0 \phi_0(\vec{r}),
\label{schrodingercond}
\end{equation}
where
the condensate
density is given by $n_0(\vec{r})= N_0 |\phi_0(\vec{r})|^2$ and
$N_0$\ is the total number of particles in the condensate.
The non-condensate density is $n'(\vec{r})
= \sum_{j \neq 0} |\phi_j(\vec{r})|^2 N(\hbar \omega_j)$, and the total density
is $n(\vec{r}) = n_0(\vec{r}) + n'(\vec{r})$. The free energy in 
this case is according to Eq.~(\ref{free10}) given by 
\begin{equation}
F = k_B T \sum_{j \neq 0}
\ln {\left( 1 - \exp{(-\beta \hbar \omega_j ) } \right) } 
- \int d\vec{r}\ [ n^2(\vec{r}) - \frac{1}{2} n_0^2(\vec{r})] T^{2B} 
+ \mu \int d\vec{r}\ n(\vec{r}).
\label{vrijeenergiecond}
\end{equation}
Note that for $T=0$, i.e.\ all particles in the groundstate, 
Eq.~(\ref{schrodingercond})
corresponds to the non-linear Schr\"odinger equation (NLSE), studied
for example by Ruprecht {\it et al.} \cite{keith}\ and first
derived by Goldman {\it et al.} \cite{tony}\ in their pioneering work
on spin-polarized atomic hydrogen.
In addition, note that Bergeman \cite{bergeman} in his 
analysis uses Eq.~(\ref{schrodinger}) with $T^{2B}$\ replaced by
$T^{2B}/2$\ for both the condensate and the excited state wave functions,
which corresponds to the Hartree approximation.
Although this gives correct results at zero temperature,
this is no longer true for non-zero temperatures because it does not
properly take into account the mean-field interactions due to the
non-condensate part of the gas. 
 
\section{RESULTS}
\label{resultaat}
At this point we have all tools available to 
study the stability of atomic $^7$Li for temperatures obeying 
$na\Lambda_{th}^2 \ll 1$.  
As mentioned previously, this is done numerically by solving 
Eq.~(\ref{schrodinger}) and Eq.~(\ref{schrodingercond})
selfconsistently with the total particle
density $n(\vec{r})$. The gas is assumed to be 
confined by an isotropic harmonic oscillator potential 
\[
V_{ext}(\vec{r}) = \frac{1}{2} m \omega^2 r^2,
\]
where we take for $\omega$\ the `average' 
$(\omega_x \omega_y \omega_z)^{1/3} $\ 
of the (non-isotropic) trap frequencies used in the
Rice experiment \cite{rice}. This results in an energy splitting of
$\hbar \omega /k_B = 7.1$nK. Due to this simplification the
density profile of the gas will depend only on the distance
$r$\ from the center of the trap. The s-wave scattering length of
$^7$Li is $a= -27.3 a_0$, where $a_0$\ is the Bohr radius \cite{randy3}.
We first consider the case $T=0$\ and subsequently present 
results for non-zero temperatures. 

\subsection{The case ${\bf T = 0}$}
\label{Tnul}
In this section all particles are considered to be in the condensate, which is
the case at zero temperature. 
This has already been subject to extensive research 
of several other papers, see for instance Refs.~\cite{keith}\ and \cite{dodd}. 
For a fixed number of condensate particles $N_0$, the lowest
energy eigenvalue and wave function of the Schr\"odinger type 
equation (\ref{schrodingercond}) 
is solved by a numerical integration and the density distribution
$n(r) = n_0(r)$, the chemical potential $\mu$\ and the free energy $F$\ 
are determined from this solution. 

In Fig.~\ref{fig1} we plot first of all $\mu$\ as a function of $N_0$. 
If there are only a
few particles in the condensate, $\mu$ is seen to be equal to 
$\frac{3}{2} \hbar \omega$, i.e.\ the groundstate energy of a particle
in a harmonic oscillator. However, as $N_0$ increases, the effective
potential $V_{eff}(r) = \frac{1}{2} m \omega^2 r^2 + n(r) T^{2B}$,
grows deeper and deeper in a small range around the center of the
trap since $T^{2B}$ is negative, which pulls the particles more and 
more to the center of the trap. This decreases the value
of the ground-state energy and consequently also the chemical potential. 
As can be seen from the
figure, for $N_0 > 1241$, 
a solution cannot be found anymore, indicating
that the condensate becomes unstable. The maximum number
of $N_{0,max} = 1241$\ corresponds well with the condition
$N_{0,max} \simeq 0.405\ l/|a|$\ found by Ruprecht {\it et al.}\ for the
appropriate trap parameters \cite{keith}. 

The free energy given by Eq.~(\ref{vrijeenergiecond}) is plotted
as a function of $N_0$ and as a function of the  
central density $n_c$ of the gas in Figs.~\ref{fig2} and 
\ref{fig3} respectively. Notice that the derivative of the free energy with
respect to the number of condensate particles exactly reproduces the
chemical potential, i.e. $\mu = \partial F / \partial N$, showing the
consistency of our calculations. 
From Fig.~\ref{fig3} it follows that the free energy as a function
of the central density approaches a constant (maximum) value. 
This is also shown in the inset of this figure.
So, as anticipated in Sec.~\ref{stab}, the instability 
appears at an extremum of the free energy as a function of the
central density of the gas cloud. If the density increases further,
the gas will collapse to a dense state. With the theory presented above
we clearly cannot describe
the gas beyond this point, for we would need a theory that can describe 
the system also at high densities. 

\subsection{The case ${\bf T \neq 0}$}
\label{Teindig}

For non-zero temperatures the particles
in the gas occupy the harmonic oscillator states in Eq.~(\ref{schrodinger})
according to the normal Bose distribution at a given chemical potential.
However, if $\mu$\ is increased from $-\infty$\ to some value
below $\frac{3}{2}\hbar \omega$, the number of particles in the 
ground state starts to increase dramatically, and we can only 
put more particles in the system by forcing them into the
ground state. At this point, $\hbar \omega_0$\ equals zero 
and the gas consists of a condensate part with density
$n_0(r) = N_0 |\phi_0(r)|^2$, as well as a non-condensate part
with density $n'(r) =\sum_{j\neq 0} N(\hbar \omega_j) |\phi_j(r)|^2$.
So, the calculation of the chemical potential and the free energy now 
consists of two parts which correspond to the use of the unrestricted
and restricted grand-canonical ensemble respectively. 
In the first part we increase $\mu$\ from $-\infty$\ to
some maximum value $\mu_{max}$, above which there are no longer
any solutions. In the second part we on the other hand increase the
number $N_0$\ of particles in the ground state and then
determine $\mu$\ from the ground-state energy 
of the trap using that $\hbar \omega_0 = 0$. Note that in this second part
the chemical potential decreases again, because the increasing density
of the condensate lowers the ground-state energy. This behaviour of the
chemical potential explains why no solutions with $\mu > \mu_{max}$\ could 
be found in the calculation with the unrestricted grand-canonical ensemble. 
The two parts of the calculation
join smoothly together within an error of the order of $1\%$\ at $\mu_{max}$\ 
where the condensate fraction is of the order of $5\%$.

At non-zero temperatures the
density profiles of the gas are determined by calculating
all energy levels and corresponding
eigenfunctions up to $10k_BT$. 
Since the ideal 3D harmonic oscillator energy levels are given by
$\varepsilon_{n,l} = (2n+l+3/2)\hbar \omega$\ where $n$\ and $l$\
are integers, this corresponds to taking as many 
as $(1/2*10k_BT/\hbar\omega)^2$\ levels into account. Clearly, 
this number increases rapidly as a function of temperature.
When the number of particles in the groundstate is small (typically
corresponding to $\mu \stackrel{<}{\sim} -\hbar \omega$), 
Eq.~(\ref{schrodinger})
was used to calculate all wave functions and Eq.~(\ref{vrijeenergie}) to
calculate the free energy. For larger values of $\mu$, the groundstate
was determined by Eq.~(\ref{schrodingercond}) and the free energy by
Eq.~(\ref{vrijeenergiecond}). To check that our results are consistent,
we first compare in Fig.~\ref{fig40}\ the density profile 
above the critical temperature with the prediction of the
local-density approximation, i.e.\ with 
\[ 
n(r) = 1/\Lambda_{th}^3\ g_{3/2}(\exp{[\beta(\mu-V(r)-2n(r)T^{2B})]}),  
\] 
The agreement is good for large and negative
$\mu$, but for $\mu \simeq - \hbar \omega$, a deviation
becomes visible around the center of the trap, indicating that
the critical temperature is approached.

In Figs.~\ref{fig5} and \ref{fig6} 
the free energy is plotted as a function of 
the (total) central density of the gas at
$T=50\mbox{nK}$\ and $T=100\mbox{nK}$\ respectively. We again checked that
$\mu = \partial F / \partial N$\ as required. 
The inset in both figures shows a magnification of the local minimum
in the free energy curve where $\mu$\ goes through zero. This is not
a point of instability of the system because small variations
in the central density do not conserve the total number
of particles. The point of instability occurs again 
where the free energy approaches a local maximum.
Note that the maximum densities in the center of the trap that can be
obtained are orders of magnitude higher then those for a 
homogeneous trap. In the homogeneous case, collapse occurs already at
densities smaller than $ n_{BEC} = \zeta (3/2) / \Lambda_{th}^3 \simeq
2.612 / \Lambda_{th}^3$.
For $T=50 \mbox{nK}$\ this corresponds to $n_{BEC} = 1.07 \cdot 
10^{11} $cm$^{-3}$, and for $T=100\mbox{nK}$\ we have $n_{BEC} = 3.04 \cdot
10^{11}$cm$^{-3}$. Note furthermore that the interaction 
term $[2n(r) - n_0(r)] T^{2B}$\
in the effective potential $V_{eff}(r)$\ of the groundstate 
becomes in the center of the trap as large as $n_c T^{2B} \simeq 
- 5\hbar \omega$.

Next we take a closer look at the point of instability of the system
when the temperature increases. As was pointed out before however,
the number of harmonic oscillator states which have to be taken into account 
to calculate the density acurately increases very rapidly
and thus slows down the calculation considerably. To avoid this, we make 
for temperatures higher than $150$nK use of the fact that 
only the lowest wave functions of the harmonic oscillator 
states are influenced by the interaction term 
$2n(r)T^{2B}$\ or $[2n(r)-n_0(r)]T^{2B}$\ for the ground state,
and the wave functions of the higher states are unaffected, although their
occupation numbers change, due to the fact that $\mu$\ equals
the ground-state energy if there is a Bose condensate present.

In Fig.~\ref{fig7} we plot at the point of instability
the normal density of the gas in the center of the trap 
$n'(0)$\ as a function of temperature (dashed
line) and compare this with the density 
$n_{BEC}\simeq 2.612/\Lambda^3_{th}$\ (solid line)
required for BEC in the homogeneous case. In the inset of the
same figure, the number of noncondensed particles as a function of
temperature is plotted (dashed line), and this is compared with 
the usual criterion for the onset of BEC in a noninteracting gas, i.e.\  
$N_{BEC} = \zeta (3) (k_BT/\hbar \omega)^3 
\simeq 1.202 (k_B T / \hbar \omega )^3$ (solid line).
As expected, the consequence of the attractive interatomic interaction
is that the non-condensed particles are pulled towards the center of the trap. 
The solid line in Fig.~\ref{fig8} shows the maximum number of 
condensate particles as a function of temperature. Clearly, the 
occupation of the condensate 
at the point where the gas becomes unstable decreases when 
the temperature increases. 
In Ref. \cite{stoof} it was argued that an increase in temperature
would {\it not} lead to a decrease in the maximum number
of condensate particles since the non-condensed
density is approximately constant over the extend of the condensate
wave function and therefore only shifts the effective potential
$V_{eff}(r)$\ by a constant. If this is true, the observed decrease 
can only be explained by the non-condensed part of the gas becoming
unstable before the condensate holds the maximum number of particles.    
This might also be physically reasonable because 
the contribution of the normal part of the gas to the total
density increases everywhere and especially around the center of the
trap.

In view of this we want to try to answer the question whether
it is the condensate or (as in the homogeneous case)
the non-condensed part of the gas which causes instability. 
Of course, for zero temperature the condensate becomes unstable, and we expect 
the same for such low temperatures that the non-condensed part is
only a small fraction of the gas. For higher temperatures this might
change since the number of non-condensed particles increases very rapidly
as a function of temperature. 
To analyse this issue, 
we calculated in the temperature interval $0 \leq T \leq 400$nK 
the density profiles of the condensate $n_0(r)$\ and the
normal part of the gas $n'(r)$\ at the spinodal point.
The condensate fraction $N_0/N$\ decreases
from 1 for $T=0$\ to about 0.005 for $T=400$nK.
Subsequently, we try to add particles to the condensate,
and try to find a new solution to the non-linear Schr\"odinger equation
Eq.~(\ref{schrodingercond}) for the increased number of condensate particles,
{\it while keeping the non-condensate density profile fixed.} 

The results are also plotted in Fig.~\ref{fig8}: The
dots in this figure denote the maximum number of particles
that can be in the condensate given the non-condensate density
at the temperature of interest.  
For temperatures $T\leq 50$nK, the system becomes already
unstable if only one particle is added to the condensate, 
from which we draw the conclusion that
at this temperature it is still the condensate which renders the instability.
For higher temperatures, it is possible to add a few particles 
to the condensate, but this appears te be the result of 
numerical unaccuracies of our calculation. We thus conclude that
the condensate is unstable at the point of instability of the whole system. 

Because of this result, we suspect that the simple argument that 
the maximum number of condensate particles remains constant
because the non-condensed part of the gas is approximately constant
over the extend of the condensate wavefunction, may not 
be sufficiently accurate 
in this system. The only difference in the condensate wave function
at different temperatures arises due to the contribution of the
term $2n'(r)T^{2B}$\ to the effective potential. In Fig.~\ref{fig9}
we plotted for $T=0$\ (and consequently $n'(r)=0$) (solid lines), 
and for $T=300$nK (dashed lines) the effective potential $V_{eff}(r)
= 1/2 m\omega^2r^2 + 2 n'(r) T^{2B}$\ and the corresponding condensate 
densities $n_0(r)$. 
When $V_{eff}(r)$\ for $T=300$nK is shifted upward such that
the zero of both potentials coincide, it is clear that 
the normal part of the gas effectively increases the oscillator strength
of the trap potential when the temperature increases. 
Since the maximum condensate size $N_{0,max} \propto l/|a| \propto
1/\sqrt{\omega}$, an increase in the effective oscillator strength felt
by the condensate causes a decrease in the maximum condensate size.
A measure for the deviation of the effective oscillator strength
from the original strength $\omega$\ is given by the
expression  
\[
\omega^2_{eff} = \omega^2 \frac{\int d\vec{r}\ 
(V_{ext}(r) + 2[n'(r)-n'(0)]T^{2B})n_0(r)}{\int d\vec{r}\ V_{ext}(r) n_0(r)},
\]
and for non-zero temperature we thus estimate that the maximum number
of condensate particles is given by 
\[
N_{0,max}(T)=N_{0,max}(0) \sqrt{\frac{\omega}{\omega_{eff}(T)}}.
\]
For $T=300$nK this amounts to $N_{0,max}(300) = 1174.5$, which corresponds
rather accurately with the value 1173 from our full calculation. For some
other temperatures the maximum occupation of the
condensate determined in this way
is denoted in Fig.~\ref{fig8}\ by the open circles.  
We can conclude that 
the growth of the normal part of the gas occurs
at the expense of the condensate when temperature increases.

\section{CONCLUSION}
\label{conclusie}
We performed a numerical calculation to study the stability of a
Bose condensate in a trapped gas of $^7$Li atoms at 
zero and non-zero temperatures. This was done by determining
all quantum states for particles in a harmonic oscillator trap and
interacting via two-body scattering. The proposed criterion that the gas becomes 
mechanically unstable when the free energy of the system as a function
of the central density of the gas approaches a maximum value,
is confirmed by the calculations. 

For zero temperature, the maximum number of condensate particles
is in agreement with previous calculations, and for non-zero 
temperature this number decreases considerably.
This is due to the fact that the condensate experiences an
effective oscillator strength due to the presence of the
non-condensed part of the gas. This effective potential increases
as temperature increases and therefore results in a decrease
of the maximum occupation of the condensate. For the temperature interval
$0 \leq T \leq 400$\ the condensed part of the gas renders the instability
at the spinodal point, so
in contrast to the homogeneous case, the normal part of the
gas remains stable against density fluctuations.

Furthermore, from the results in Sec.~\ref{Teindig}\ and the discussion
in Sec.~\ref{bec}, it can be concluded that at low temperatures 
it seems necessary to include also many-body effects in e.g.\ 
the scattering length, since the average interaction $n T^{2B}$\ becomes 
substantially larger than the energy splitting $\hbar \omega$. 
To do so appears to be an important challenge for
the future which is not only difficult in practice but even in principle
due to the presence of infra-red divergences in the theory of the
dilute Bose gas \cite{russen}. Closely related to this issue is
the effect of the BCS transition on the properties of the gas, which
still needs to be incorporated in the numerical calculations. 
Once the experiments enter into this low temperature regime
where $na\Lambda^2_{th} \gg 1$, it should be interesting to compare
the experimental data with the mean-field analysis presented here,
and to see if possible deviations can be understood by the above
mentioned corrections. This is of course not only true for $^7$Li,
but also for any other atomic species with a negative scattering length
such as $^{85}$Rb and $^{123}$Cs.
  
\section*{ACKNOWLEDGMENTS}
We would like to thank Chris Pethick for pointing out to us the analogy
between the metastable Bose condensate and compact objects and
the stability criteria of the latter system. Furthermore
we thank Randy Hulet and Keith Burnett for various enlightening 
discussions. 
 
\section*{APPENDIX}
\setcounter{equation}{0}
\renewcommand{\theequation}{A.\arabic{equation}}

To calculate the minimum of the grand-canonical potential $\Omega$\ of
Eq.~(\ref{omega}), it is convenient in the following to introduce
a compact notation for the inner product of two states.
From Eq.~(\ref{normering}), it is found that the normalization can be rewritten
as
\begin{eqnarray}
1 & = &
\int d\vec{r}\  \left( |u_j(\vec{r})|^2 - |v_j(\vec{r})|^2 \right)
  \nonumber \\ & =  &
\int d\vec{r} \ \left( \begin{array}{ccc}
        u_j^*(\vec{r}) & , & - v_j^*(\vec{r}) \end{array} \right)
\left( \begin{array}{cc} 1 & 0 \\ 0 & 1 \end{array} \right)
\left( \begin{array}{c} u_j(\vec{r}) \\ v_j(\vec{r}) \end{array} \right) 
\nonumber \\ 
& \equiv & ( j |1| j ),
\label{inner}
\end{eqnarray}
which thus defines our convention for the inner product.

We start with the derivative of $\Omega$\ with respect to $\hbar\Sigma$. 
It is given by
\begin{eqnarray}
\frac{\partial \Omega}{\partial \hbar \Sigma} & = & 
\frac{\partial E_g}{\partial \hbar \Sigma} + 
\sum_j N(\hbar\omega_j) \frac{\partial \hbar \omega_j}{\partial \hbar\Sigma}
+ \int d\vec{r} \left\{  
2 T^{2B} \langle \psi^{\dag}(\vec{r}) \psi(\vec{r}) \rangle_t
\frac{ \partial \langle \psi^{\dag}(\vec{r}) \psi(\vec{r}) \rangle_t}{\partial
\hbar\Sigma} \right. \nonumber \\ & & 
+  \frac{1}{2} V_0 
\frac{ \partial \langle \psi^{\dag}(\vec{r}) \psi^{\dag}(\vec{r}) \rangle_t
}{\partial \hbar\Sigma} \langle \psi(\vec{r}) \psi(\vec{r}) \rangle_t
+ \frac{1}{2} V_0 
\frac{ \partial \langle \psi(\vec{r}) \psi(\vec{r}) \rangle_t
}{\partial \hbar\Sigma} \langle \psi^{\dag}(\vec{r}) \psi^{\dag}(\vec{r}) \rangle_t
  - \langle \psi^{\dag}(\vec{r}) \psi(\vec{r}) \rangle_t \nonumber \\ 
& & \left. -\hbar\Sigma
\frac{ \partial \langle \psi^{\dag}(\vec{r}) \psi(\vec{r}) \rangle_t}{\partial
\hbar\Sigma}  
-  \frac{1}{2} \Delta_0(\vec{r}) 
\frac{ \partial \langle \psi^{\dag}(\vec{r}) \psi^{\dag}(\vec{r}) \rangle_t
}{\partial \hbar\Sigma}
- \frac{1}{2} \Delta_0^*(\vec{r}) 
\frac{ \partial \langle \psi(\vec{r}) \psi(\vec{r}) \rangle_t
}{\partial \hbar\Sigma} \right\}. 
\end{eqnarray}
Equating the whole expression to zero, and
grouping together the terms proportional to the derivative of an
expectation value with respect to $\hbar \Sigma$, the solution is 
seen to be given by Eq.~(\ref{sigma}) and Eq.~(\ref{gap}), if we can 
prove that 
\begin{equation} 
\frac{\partial E_g}{\partial \hbar \Sigma} + 
\sum_j N(\hbar\omega_j) \frac{\partial \hbar \omega_j}{\partial \hbar\Sigma}
 - \int d\vec{r} \ \langle \psi^{\dag}(\vec{r}) \psi(\vec{r}) \rangle_t = 0. 
\label{nul}
\end{equation}

This is most easily achieved by assuming that all functions $u_j(\vec{r})$,
$v_j(\vec{r})$\ and $\Delta_0(\vec{r})$\ are real. In that case,
the first term on the left reduces to
\begin{equation}
\frac{\partial E_g}{\partial \hbar \Sigma}   
 = - \sum_j \left\{ \int d\vec{r}\ 
\frac{\partial \hbar\omega_j}{\partial\hbar\Sigma} v_j^2(\vec{r}) 
+ \hbar\omega_j \frac{\partial}{\partial \hbar\Sigma} 
\int d\vec{r} v_j^2(\vec{r}) \right\}.
\label{egap}
\end{equation}
The derivative $\partial \hbar \omega_j / \partial \hbar \Sigma$\ 
can be calculated by perturbing the Hamiltonian according
to
\[
\delta H = \left( \begin{array}{cc}
                \delta \hbar \Sigma & 0 \\
                0 & - \delta \hbar \Sigma \end{array} \right).
\]
The energy shift in $\hbar\omega_j$\ is then, to first order, given by
\[
\delta \hbar \omega_j = ( j|\delta H|j ) = 
\int d\vec{r}\ (u_j^2(\vec{r}) + v_j^2(\vec{r})) \delta \hbar\Sigma,
\]
and therefore
\begin{equation}
\frac{\partial \hbar \omega_j}{\partial \hbar \Sigma} =
\int d\vec{r}\ \left( u_j^2(\vec{r}) + v_j^2(\vec{r}) \right).
\label{afgeleide}
\end{equation}
Using furthermore that $\int d\vec{r} \ \left( u_j^2(\vec{r}) -
v_j^2(\vec{r}) \right)  = 1$, it is found that
\begin{equation}
\int d\vec{r} \ v_j^2(\vec{r}) = \frac{1}{2} \left( 
\frac{\partial \hbar \omega_j}{\partial \hbar \Sigma} -1 \right),
\label{v2}
\end{equation}
so Eq.~(\ref{egap}) can be rewritten as
\begin{eqnarray}
\frac{\partial E_g}{\partial \hbar \Sigma} & = &
- \frac{1}{2} \sum_j\ \left( 
\frac{\partial \hbar \omega_j}{\partial \hbar \Sigma} -1 \right)
\frac{\partial \hbar \omega_j}{\partial \hbar \Sigma} 
- \frac{1}{2} \sum_j \hbar \omega_j \frac{\partial^2
\hbar \omega_j}{\partial \hbar \Sigma^2} \nonumber \\
& = & \sum_j  \int d\vec{r} \ v_j^2(\vec{r})  + \frac{1}{2}
\sum_j \left( 1 - \left( 
\frac{\partial \hbar \omega_j}{\partial \hbar \Sigma} \right)^2 -
\hbar \omega_j
\frac{\partial^2 \hbar \omega_j}{\partial \hbar \Sigma^2} \right).
\label{egap2}
\end{eqnarray}   

The factor between brackets is zero. This can be seen by first
applying a Taylor expansion to show that the energy change 
$\delta \hbar \omega_j$\ due to the shift $\hbar \Sigma
+ \delta \hbar \Sigma$\ can be written as 
\[
\delta \hbar \omega_j =  
\frac{\partial \hbar \omega_j}{\partial \hbar \Sigma} \delta\hbar \Sigma
+ \frac{1}{2} 
\frac{\partial^2 \hbar \omega_j}{\partial \hbar \Sigma^2} 
(\delta\hbar \Sigma)^2 + {\cal{O}}\left( (\delta \hbar \Sigma)^3 \right).
\]
Therefore, the last term between the brackets in Eq.(\ref{egap2}) 
can be identified with twice the second-order
energy shift. Making then use of the standard expression for this 
second-order energy shift
\begin{equation}
\delta \hbar \omega_j = (j|\delta H | j ) + \sum_{i \neq j}
\frac{( j |\delta H |i )( i|\delta H |j)}{
\hbar \omega_j -\hbar \omega_i} + .... \ ,
\label{energieshift}
\end{equation}
which still holds with our new definition of the inner product, we obtain
\begin{eqnarray*}
\sum_j \left( 1 - \left( 
\frac{\partial \hbar \omega_j}{\partial \hbar \Sigma} \right)^2 -
\hbar \omega_j
\frac{\partial^2 \hbar \omega_j}{\partial \hbar \Sigma^2} \right) & = &
 \\ 
\sum_j \left( 1 - ( j| \frac{\partial H}{\partial \hbar \Sigma} |j
)( j| \frac{\partial H}{\partial \hbar \Sigma} |j) 
- 2 \hbar \omega_j \sum_{i \neq j} \frac{
( j| \frac{\partial H}{\partial \hbar \Sigma} |i) 
( i| \frac{\partial H}{\partial \hbar \Sigma} |j )}{
\hbar \omega_j - \hbar\omega_i} \right) & = & \\
 \sum_j \left( 1 - ( j| \left( \frac{\partial H}{\partial \hbar \Sigma}
\right)^2 |j ) \right)& = & 0, 
\end{eqnarray*}
where the third line follows from the completeness of the eigenstates
of the Bogoliubov-DeGennes equation and 
from the fact that if we write $2 \hbar \omega_j = (\hbar \omega_j
-\hbar \omega_i) + (\hbar \omega_j + \hbar \omega_i)$\ only
the antisymmetric part contributes. Finally, we also need that 
\[
\frac{\partial H}{\partial \hbar \Sigma} = \left( \begin{array}{cc}
1 & 0 \\0 & -1 \end{array} \right).
\] 
Collecting all terms together we thus indeed find that (see Eq.~(\ref{nul}))
\begin{eqnarray}
\frac{\partial E_g}{\partial \hbar \Sigma} + 
\sum_j N(\hbar\omega_j) \frac{\partial \hbar \omega_j}{\partial \hbar\Sigma}
 - \int d\vec{r} \ \langle \psi^{\dag}(\vec{r}) \psi(\vec{r}) \rangle_t & = &
\nonumber \\
\int \ d\vec{r} \sum_j \left\{ v_j^2(\vec{r}) + N(\hbar\omega_j)
( u_j^2(\vec{r}) + v_j^2(\vec{r})) \right\} - 
\int d\vec{r} \ \langle \psi^{\dag}(\vec{r})
\psi(\vec{r}) \rangle_t  & = & 0.
\end{eqnarray} 

At the same time the derivative of $\Omega$\ with respect to
$\Delta_0$\ must be zero. That this is also the case can be shown by a
similar calculation. Assuming $\Delta_0$, $u_j$\ and $v_j$\ to be real, 
the equation $ \partial \Omega / \partial \Delta_0 = 0$
\ reduces to
\begin{equation} 
\frac{\partial E_g}{\partial \Delta_0} + \sum_j N(\hbar\omega_j)
\frac{\partial \hbar\omega_j}{\partial \Delta_0} - \frac{1}{2}
\langle \psi \psi \rangle_t - \frac{1}{2} \langle \psi^{\dag} \psi^{\dag}
\rangle_t = 0
\label{upsie2}
\end{equation}
if the functions $\hbar\Sigma$\ and $\Delta_0$\ are given by Eqs.~(\ref{sigma})
and (\ref{gap}). Using Eq.~(\ref{v2}), the 
first term on the left-hand side is equal to
\begin{eqnarray}
\frac{\partial E_g}{\partial \Delta_0} & = & - \sum_j \left\{ \frac{\partial
\hbar \omega_j}{\partial \Delta_0} \int d\vec{r}\ v_j^2(\vec{r})
+ \hbar \omega_j \frac{\partial }{ \partial \Delta_0} 
\int d\vec{r}\ v_j^2(\vec{r}) \right\} \nonumber \\ 
& = & -\frac{1}{2} \sum_j \left\{ \frac{\partial \hbar\omega_j}{\partial
\Delta_0} \right( \left. 
\frac{\partial \hbar \omega_j}{\partial \hbar \Sigma} - 1
\right) + \hbar\omega_j \left. \frac{\partial^2 \hbar \omega_j}{\partial
\Delta_0 \partial \hbar \Sigma} \right\}.
\label{upsie}
\end{eqnarray} 
As in the previous calculation,
the first and second derivatives of $\hbar\omega_j$\ with respect to
$\Delta_0$\ and $\hbar \Sigma$\ can be calculated by perturbing the
Hamiltonian according to $H = H_0 + \delta H$, with
\[
\delta H =
\left( \begin{array}{cc}
\delta \hbar \Sigma & \delta \Delta_0 \\ -\delta \Delta_0 & 
- \delta \hbar \Sigma \end{array} \right).
\]
The energy levels then shift to second order according to
\begin{eqnarray}
\delta \hbar \omega_j & = & \frac{\partial \hbar\omega_j}{\partial \hbar \Sigma}
\delta\hbar \Sigma + \frac{\partial \hbar\omega_j}{\partial \Delta_0}
\delta \Delta_0
+ \nonumber \\ & & + \frac{1}{2} \left( 
\frac{\partial^2 \hbar\omega_j}{\partial \hbar \Sigma^2} \delta\hbar\Sigma^2+
2 \frac{\partial^2 \hbar\omega_j}{\partial \hbar \Sigma \partial
\Delta_0 } \delta\hbar\Sigma \delta \Delta_0 +
\frac{\partial^2 \hbar\omega_j}{\partial \Delta_0^2} \delta \Delta_0^2 \right).
\end{eqnarray}
Comparing this again with the perturbative expression for the energy shift
Eq.~(\ref{energieshift}),
we find that 
\begin{equation}
\frac{\partial \hbar \omega_j}{\partial \Delta_0} = \frac{
( j|\delta H | j)}{\delta \Delta_0} =( 
j| \frac{\partial H}{\partial \Delta_0} |j ) = 2\int d\vec{r}
\ u_j(\vec{r}) v_j(\vec{r}).
\end{equation}
Moreover, the second order derivative in Eq.~(\ref{upsie}) can be written as
\begin{eqnarray}
\frac{\partial^2 \hbar \omega_j}{\partial \Delta_0 \partial \hbar \Sigma} & = &
\sum_{i \neq j} \frac{
( j|\frac{\partial H}{\partial \Delta_0}|i)( i|\frac{
\partial H}{\partial \hbar \Sigma} | j ) + ( j| \frac{\partial
H}{\partial \hbar\Sigma}|i)( i|\frac{\partial H}{\partial
\Delta_0} |j )}{\hbar \omega_j - \hbar \omega_i}.
\end{eqnarray}
Therefore Eq.~(\ref{upsie}) becomes
\begin{eqnarray}
\frac{\partial E_g}{\partial \Delta_0} & = &
\sum_j \left\{ \int d\vec{r} \ u_j(\vec{r}) v_j(\vec{r})  \ -\frac{1}{2}
( j| \frac{\partial H}{\partial \Delta_0} |j )( j|
\frac{\partial H}{\partial \hbar \Sigma } | j )  \right.
\nonumber \\ & & - \left. \frac{1}{2} \sum_{i \neq j}
\frac{\hbar\omega_j }{ \hbar \omega_j - \hbar \omega_i} \times  \left[ 
( j|\frac{\partial H}{\partial \Delta_0}|i)( i|\frac{
\partial H}{\partial \hbar \Sigma} | j ) + ( j| \frac{\partial
H}{\partial \hbar\Sigma}|i)( i|\frac{\partial H}{\partial
\Delta_0} |j ) \right] \ \right\} \nonumber \\
& = & \sum_j \int d\vec{r}\ u_j(\vec{r}) v_j(\vec{r}) \ -\frac{1}{4}
\sum_{j} \left[ 
( j|\frac{\partial H}{\partial \Delta_0}\frac{
\partial H}{\partial \hbar \Sigma} | j ) + ( j| \frac{\partial
H}{\partial \hbar\Sigma} \frac{\partial H}{\partial
\Delta_0} |j ) \right] \nonumber \\
& = &  \sum_j \int d\vec{r}\ u_j(\vec{r}) v_j(\vec{r}), 
\end{eqnarray}
since 
\[
\frac{\partial H}{\partial \Delta_0} = \left(
\begin{array}{cc} 0 & 1 \\ -1 & 0 \end{array} \right) \ \ \mbox{and}
\ \ \frac{\partial H}{\partial \hbar \Sigma} = \left(
\begin{array}{cc} 1 & 0 \\ 0 & -1  \end{array} \right).
\]
Combining all results, Eq.~(\ref{upsie2}) becomes
\begin{eqnarray}
\frac{\partial E_g}{\partial \Delta_0} + \sum_j N(\hbar\omega_j)
\frac{\partial \hbar\omega_j}{\partial \Delta_0} - \frac{1}{2}
\langle \psi \psi \rangle_t - \frac{1}{2} \langle \psi^{\dag} \psi^{\dag}
\rangle_t & = & \nonumber \\
\sum_j \int d\vec{r}\ u_j(\vec{r}) v_j(\vec{r}) (1 + 2 N(\hbar \omega_j))
- \frac{1}{2}\langle  \psi \psi \rangle_t - \frac{1}{2}
\langle \psi^{\dag} \psi^{\dag} \rangle_t & = & 0
\end{eqnarray}
which completes the proof.

To gain even more confidence in our expression for $\Omega$, it is useful to 
see whether 
\begin{equation}
N= \int d\vec{r}\ n(\vec{r}) = 
\int d\vec{r}\ \langle \psi^{\dag}(\vec{r}) \psi(\vec{r}) \rangle_t =
- \left. \frac{\partial \Omega}{\partial \mu} \right|_{\hbar \Sigma, 
\Delta_0^{\dag},\Delta_0}
\end{equation}
is satisfied. To prove this, we
first notice that $n^2(\vec{r}) = (\hbar \Sigma (\vec{r})/2 T^{2B})^2$\ and
that according to the matrix form of the Hamiltonian
(cf. Eq.~(\ref{matrix}))
\begin{equation}
\frac{\partial \hbar \omega_j}{\partial \mu} = - \frac{\partial \hbar
\omega_j}{\partial \hbar \Sigma}.
\end{equation}
Therefore, we can make use of our previous results in  
Eqs.~(\ref{afgeleide}) and (\ref{egap2}) to obtain
\begin{eqnarray}
\left. \frac{\partial \Omega}{\partial \mu } \right|_{\hbar \Sigma, 
\Delta_0^{\dag}, \Delta_0} & = & \frac{\partial E_g}{\partial \mu} 
+ N(\hbar\omega_j) \frac{\partial \hbar \omega_j}{\partial \mu} 
\nonumber \\
& = & - \left\{ \frac{\partial E_g}{\partial \hbar \Sigma} +
N(\hbar\omega_j) \frac{\partial \hbar \omega_j}{\partial \hbar \Sigma} \right\}
\nonumber \\ & = &
-\sum_j \int d\vec{r}\ \left\{ (|u_j(\vec{r})|^2 + |v_j(\vec{r})|^2)N(\hbar \omega_j)
+|v_j(\vec{r})|^2 \right\} \nonumber \\ & = & - \int d\vec{r}\ n(\vec{r}),
\nonumber  
\end{eqnarray}
as desired.


\begin{figure}
\caption{Chemical potential as a function of total number
of condensate particles.}
\label{fig1}
\end{figure}

\begin{figure}
\caption{Free energy as a function of total number of condensate
particles.}
\label{fig2}
\end{figure}

\begin{figure}
\caption{Free energy as a function of the central density. The inset 
shows that the derivative of the free energy with respect to the
central density approaches zero at the point of instability and that
at this point the number of particles as a function of the central
density exhibits a maximum.}
\label{fig3}
\end{figure}

\begin{figure}
\caption{Comparison between exact density and $1/\Lambda_{th}^3\ 
g_{3/2}(\zeta)$\ above the critical temperature for (1) $\mu = - \hbar
\omega$, (2) $\mu= -10 \hbar \omega$\ and (3) $\mu = -25 \hbar \omega$.}
\label{fig40}
\end{figure}

\begin{figure}
\caption{Free energy as a function of the central density for T=50nK.}
\label{fig5}
\end{figure}

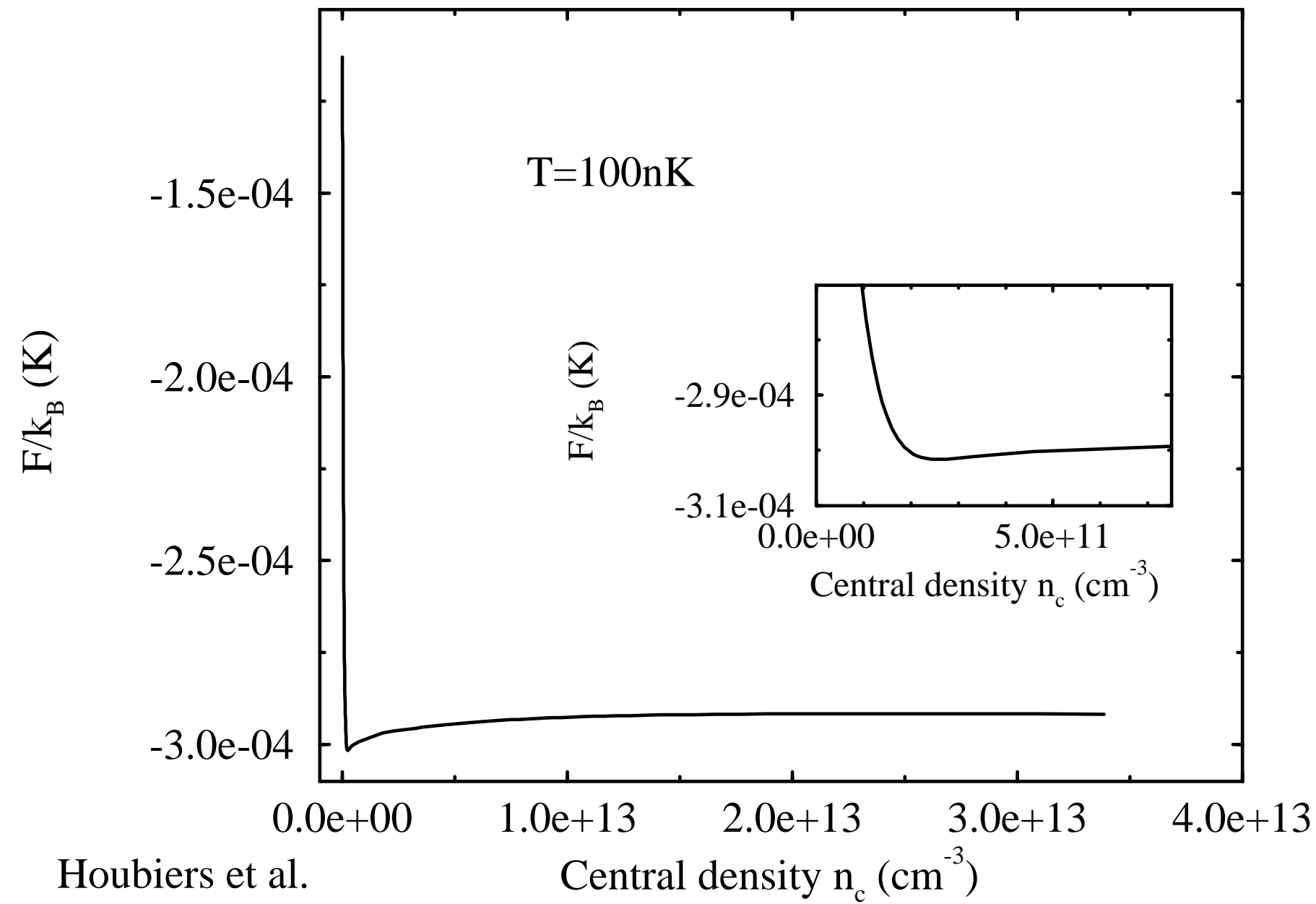
\begin{figure}
\caption{Free energy as a function of the central density for T=100nK.}
\label{fig6}
\end{figure}

\begin{figure}
\caption{Normal density $n'(0)$\ in the center of the trap at the point
of instability of the system (dashed line). The solid line 
shows $n_{BEC}=2.612/\Lambda_{th}^3$.
The inset shows the number of non-condensed 
particles as a function of temperature 
for the harmonic oscillator with (dashed) and without interaction (solid),
which is given by $N'= 1.202 (k_BT/\hbar \omega)^3$}
\label{fig7}
\end{figure}

\begin{figure}
\caption{Maximum number of condensate particles (solid line). The dots denote
the maximum condensate number at fixed normal part, and the open circles
give the maximum occupation of the condensate that can be calculated from the
effective oscillator strength due to the presence of the non-condensed
part of the gas.}
\label{fig8}
\end{figure}

\begin{figure}
\caption{Condensate density (right scale) and effective potential
(left scale) for $T=0$\ (solid lines) and $T=300$nK (dashed lines).}
\label{fig9}
\end{figure}


\begin{references}
\bibitem{JILA} M.H. Anderson, J.R. Ensher, M.R. Matthews,
               C.E. Wieman, and E.A. Cornell,
               Science {\bf 269}, 198 (1995).
\bibitem{MIT} K.B. Davis, M.O. Mewes, M.R. Andrews,
              N.J. van Druten, D.S. Durfee, D.M. Kurn, and
              W. Ketterle,
              Phys. Rev. Lett. {\bf 75}, 3969 (1995).
\bibitem{rice} C.C. Bradley, C.A. Sackett, J.J. Tollett, and
               R.G. Hulet,
               Phys. Rev. Lett. {\bf 75}, 1687 (1995).
\bibitem{henk} H.T.C. Stoof,
               Phys. Rev. A {\bf 49}, 3824 (1994).
\bibitem{randy2} R.G. Hulet, private communication.
\bibitem{keith} P.A. Ruprecht, M.J. Holland, K. Burnett, and
                M. Edwards,
                Phys. Rev. A {\bf 51}, 4704 (1995).
\bibitem{stoof} H.T.C. Stoof (unpublished).
\bibitem{michel} M. Bijlsma and H.T.C. Stoof (unpublished)
\bibitem{FenW} See for example 
               A.L. Fetter and J.D. Walecka, {\it Quantum Theory of Many
               Particle Systems} (McGraw-Hill, New York, 1971). 
\bibitem{falk} H. Falk, Am. J. Phys. {\bf 38}, 858 (1970).
\bibitem{degennes} P. de Gennes, {\it Superconductivity of metals
                and alloys} (Addison-Wesley, New York, 1966).
\bibitem{henk2} H.T.C. Stoof, M. Bijlsma and M. Houbiers (accepted
                for publication in NIST Journal of Research).
\bibitem{bergeman} T. Bergeman (unpublished).
\bibitem{tony} V.V. Goldman, I.F. Silvera and A.J. Leggett, Phys. Rev. B
               {\bf 24}, 2870 (1981). 
\bibitem{randy3} E.R.I. Abraham, W.I. McAlexander, C.A. Sackett,
                 and R.G. Hulet,
                 Phys. Rev. Lett. {\bf 74}, 1315 (1995). 
\bibitem{dodd} R.J. Dodd, M. Edwards, C.J. Williams, C.W. Clarck,
               M.J. Holland, P.A. Ruprecht and K. Burnett (unpublished);
               F. Dalfovo and S. Stringari, Phys. Rev. A {\bf 53},
                 2477 (1996).
\bibitem{shapiro} See S.L. Shapiro and S.A. Teukolsky, {\it Black Holes,
              White Dwarfs, and Neutron Stars - The physics of compact
              objects} (John Wiley \& Sons, New York, 1983) and references 
              therein.
\bibitem{russen} Yu.A. Nepomnyashchi\u{i} and 
                 A.A. Nepomnyashchi\u{i},
                 Zh. Eksp. Teor. Fiz. {\bf 75}, 976 (1978) [Sov. Phys.
                 JETP {\bf 48}, 493 (1978)].
\end{references}
\end{document}